\documentclass{pnastwo}

\usepackage{amssymb,amsfonts,amsmath,graphicx}

\def \ovii   {O\,{\sc vii}}
\def \oviii  {O\,{\sc viii}}
\def \neix   {Ne\,{\sc ix}}

\def \mgxii  {Mg\,{\sc xii}}

\def \sixiii {Si\,{\sc xiii}}

\def \cha {{\em Chandra}}
\def \xmm {{\em XMM-Newton}}
\def \hetgs {{\sc hetgs}}

\contributor{Submitted to Proceedings
of the National Academy of Sciences of the United States of America}
\url{www.pnas.org/cgi/doi/10.1073/pnas.0709640104}
\copyrightyear{2008}
\issuedate{Issue Date}
\volume{Volume}
\issuenumber{Issue Number}

\begin{document}



\title{X-ray emission processes in stars}



\author{Paola Testa\affil{1}{Smithsonian Astrophysical Observatory, MS-58,
	60 Garden st, Cambridge, MA 02138, USA}}

\contributor{Submitted to Proceedings of the National Academy of Sciences
of the United States of America}

\maketitle

\begin{article}

\begin{abstract} 
A decade of X-ray stellar observations with \cha\ and 
\xmm\ has led to significant advances in our understanding 
of the physical processes at work in hot (magnetized) plasmas in stars 
and their immediate environment, providing new perspectives and 
challenges, and in turn the need for improved models.
The wealth of high-quality stellar spectra has allowed us to investigate, 
in detail, the characteristics of the X-ray emission across the HR diagram. 
Progress has been made in addressing issues ranging from classical 
stellar activity in stars with solar-like dynamos (such as flares, 
activity cycles, spatial and thermal structuring of the X-ray emitting 
plasma, evolution of X-ray activity with age), to X-ray generating 
processes (e.g. accretion, jets, magnetically confined winds) that 
were poorly understood in the pre-\cha/\xmm\ era.
I will discuss the progress made in the study of high energy stellar 
physics and its impact in a wider astrophysical context, focusing on 
the role of spectral diagnostics now accessible. 
\end{abstract}





\dropcap{I}n this article I will discuss some recent progress in
our understanding of X-ray emission processes in stars, with
emphasis towards advances made possible by high-resolution X-ray 
spectroscopy. This article will necessarily focus on a few selected 
topics, as tremendous progress has been made in the field 
in the past decade of \cha\ and \xmm\ observations, greatly widening 
our horizons in the study of X-rays from normal stars.

More than half a century of X-ray stellar observations since the first
detection of solar X-ray emission \cite{Friedman51} have revealed the very 
rich phenomenology of physical processes at work in the outer atmosphere
of stars, and their immediate environments.
The first systematic observations of X-rays from stars with space
observatories revealed that X-ray emission is common in about all types 
of stars across the Hertzsprung-Russell (HR) diagram though with rather 
distinct characteristics for different types of stars, pointing to
different underlying production mechanisms \cite{Vaiana81}. 
Most late-type stars are X-ray sources, often highly variable, with 
levels of X-ray emission spanning more than four orders of magnitude 
and saturating at a level of fractional X-ray over bolometric luminosity 
$L_{\rm X}/L_{\rm bol} \sim 10^{-3}$. The Sun is close to the low activity 
end of the observed range of X-ray luminosity, with its 
$L_{\rm X}/L_{\rm bol}$ ranging between $\sim 10^{-7}$ and 
$\sim 10^{-6}$ during its activity cycle.
Massive stars on the other hand typically show low levels of X-ray
variability, and $L_{\rm X}/L_{\rm bol} \sim 10^{-7}$, consistent with
a scenario where X-rays are produced in shocks due to instabilities
in the radiatively driven winds (e.g., \cite{LucyWhite80,Owocki88}).
Only in a small range of spectral types, from late B to mid-A, are
stars observed to be X-ray dark or extremely weak emitters (e.g., 
\cite{Czesla07a,Schroeder07}).

Though the basic characteristics of X-ray stellar emission across the
HR diagram had been outlined already by previous X-ray observatories, 
the high sensitivity and spectral resolution of \cha\ and \xmm\ have 
provided novel diagnostics which can probe in detail the physics of hot
magnetized plasma. These physical processes are also at work in other 
very different astrophysical environments albeit on very different 
energy and temporal scales.
High-resolution spectroscopy of stars is producing significant new insights, 
for instance providing precise temperature and abundance diagnostics,
and, for the first time in the X-ray range, diagnostics
of density and optical depth.

In the following I will attempt to provide an overview of our current
understanding of the X-ray emission mechanisms in massive stars, of
the progress in our knowledge of the X-ray activity in
solar-like stars, and of selected aspects of the X-ray physics of
stars in their early evolution stages in pre-main sequence. 
In fact, X-ray stellar studies during this past decade have undergone
a shift of focus toward the early phases of stellar evolution,
and the study of the interplay between circumstellar environment and 
X-ray activity. E.\ Feigelson's article in this same issue addresses
the effects of the X-ray emission from the star on its 
circumstellar environment, on the evolution of the disk, formation 
of planets, and planetary atmospheres, which are not discussed in
this article. 

\section{X-ray emission in early-type stars: winds (and magnetic fields)}
\label{sec:hotstars}
Early findings of approximately constant $L_{\rm X}/L_{\rm bol}$
for early-type stars, and the low variability of X-ray emission, were 
well explained by a model in which X-rays originate in shocks produced by 
instabilities in the radiatively driven winds of these massive stars
(e.g.,\cite{LucyWhite80,Owocki88}).

These models yield precise predictions for the shapes and shifts of
X-ray emission lines, and models can therefore be tested in detail 
by deriving information on the line formation radius, overall wind 
properties, and absorption of overlying cool material. 
The high spectral resolution of \cha\ and \xmm, and especially the 
High Energy Transmission Grating Spectrometer (\hetgs, \cite{Canizares05}) 
onboard \cha\, have revealed 
a much more complex scenario than the standard model described above.
In particular, deviations from the standard model seem to suggest 
that magnetic fields likely play a significant role in some
early-type stars. Magnetic fields have in fact recently been detected
in a few massive stars (e.g., \cite{Donati02}) -- most likely fossil 
fields, because no dynamo mechanism of magnetic field production is 
predicted to exist for these massive stars since they lack a convective 
envelope. 

High resolution spectra of several massive stars are mostly consistent 
with the standard wind-shock model, with soft spectra, and blue-shifted, 
asymmetric and broad ($\sim 1000$~km~$s^{-1}$) emission lines: e.g.,
$\zeta$~Pup \cite{Cassinelli01}, $\zeta$~Ori \cite{Cohen06}.
Other sources, while characterized by the soft emission predicted by 
wind-shock models, have spectral line profiles  that are rather symmetric, 
unshifted and narrow with respect to model expectations: e.g., 
$\delta$~Ori \cite{Miller02}, $\sigma$~Ori \cite{Skinner08}.
Furthermore, a few sources have strong hard X-ray emission with many lines 
narrower than wind-shock model predictions: e.g., $\theta^1$~Ori~C
\cite{Gagne05}, $\tau$~Sco \cite{Cohen03}. For this last class of 
X-ray sources the presence of magnetic fields provides a plausible 
explanation for the observed deviations from the wind-shock model: 
the magnetic field can confine the wind which yields hotter plasma and 
narrower lines, as shown for instance for the case of $\theta^1$~Ori~C
by Gagn{\'e} et al.\ through detailed magneto-hydrodynamic simulations
which successfully reproduce the observed plasma temperature, $L_{\rm X}$,
and rotational modulation \cite{Gagne05}.

An important diagnostic for early-type stars is provided by the He-like 
triplets (comprising $r$ resonance, $i$ intercombination, and $f$ 
forbidden lines): the metastable upper level of the $f$ line can be
depopulated, populating the upper level of the $i$ transition, through
absorption of UV photons. Therefore, the  $f/i$ ratio depends on the 
intensity of the UV field produced by the hot photosphere, i.e.\ the 
distance from the photosphere at the location where the given lines 
form. The $f/i$ ratio is also density sensitive and 
can be expressed as $R = f/i = R_0 / [1 + \phi/\phi_c + n_e/n_c]$, 
where $\phi_c$ is a critical value of the UV intensity at the energy 
coupling the $f$ and $i$ upper levels, and $n_c$ is the density critical
value; we note however that densities are generally expected to be
below $n_c$.
The observed He-like line intensities appear to confirm the wind-shock 
model when the spatial distribution of the X-ray emitting plasma is 
properly taken into account \cite{Leutenegger06}.
However there are still unresolved issues.
For instance X-ray observations imply opacities that are low
and incompatible with 
the mass loss rates derived otherwise (see e.g., \cite{Owocki01}).

\section{Cool stars and the solar analogy}
The Sun, thanks to its proximity, is at present the only star that can 
be studied at a very high level of detail, with high spatial and 
temporal resolution, and it is usually used as a paradigm 
for the interpretation of the X-ray emission of other late-type stars. 
However, while the solar analogy certainly seems to apply to some 
extent to other cool stars, it is not yet well understood how different 
the underlying processes are in stars with significantly different 
stellar parameters and X-ray activity levels. 

\subsection{X-ray activity cycles}  
The $\sim 11$~yr cycle of activity is one of the most manifest 
characteristics of the X-ray emission of the Sun, and yet in other 
stars it is very difficult to observe. This is because it is 
intrinsically challenging to carry out regular monitoring of 
stellar X-ray emission over long enough time scales, and to 
confidently identify long term cyclic variability from short term 
variations that are  not unusual in cool stars (e.g., flares, rotational 
modulation).
Long term systematic variability similar to the Sun's cycle has now been
observed in three solar-like stars: HD~81809 (G5V, \cite{Favata08}),
61~Cyg~A (K5V, \cite{Hempelmann06}), $\alpha$~Cen~A (G2V, \cite{Ayres09}).
The existence of X-ray cycles in other stars nicely confirms the
solar-stellar analogy, and it is also potentially useful in order to better
understand the dynamo activity on the Sun, which remains a significant
challenge.

\subsection{The Sun in time} 
\begin{figure*}
\centerline{\includegraphics[width=16cm]{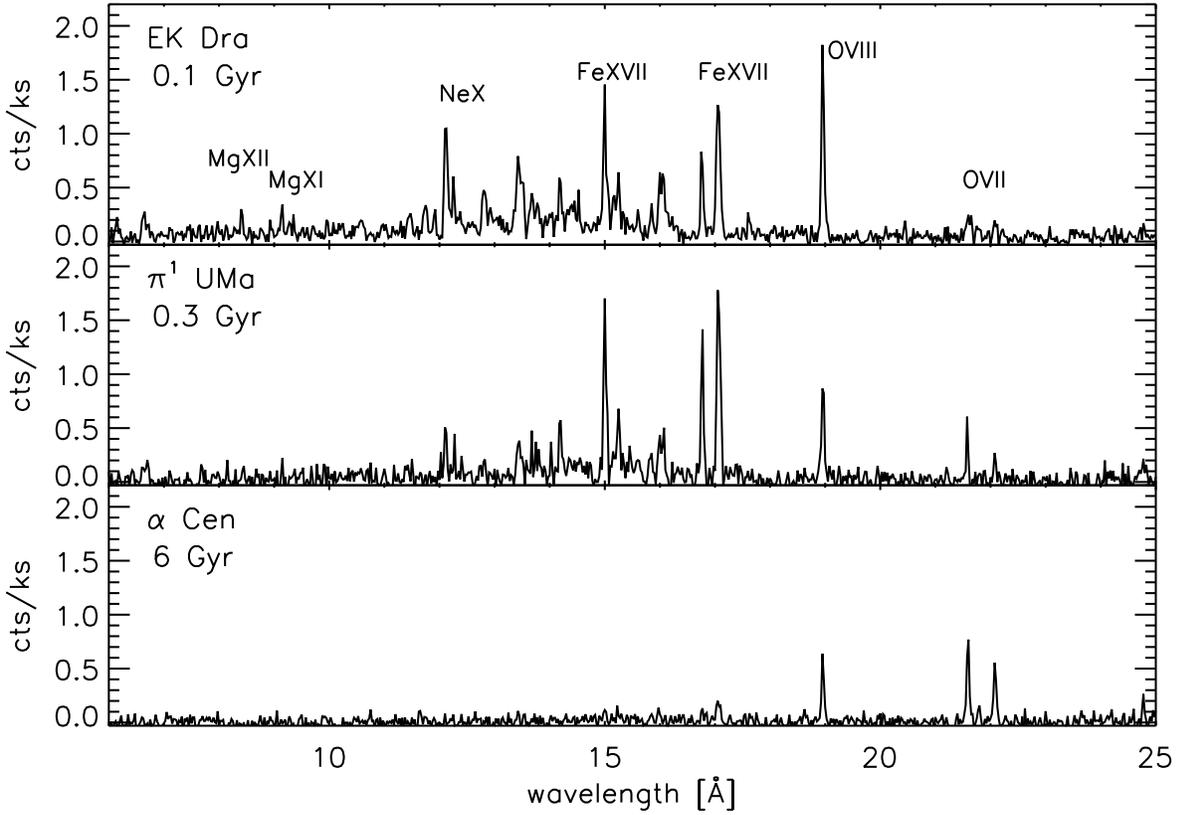}}
\caption{X-ray \cha\ Low Energy Transmission Grating (LETG)
	 spectra of solar-like stars at different
	 ages, showing the evolution of the X-ray spectrum from 
	 an age of $\sim 100$~Myr to $\sim 6$~Gyr. From top to
	 bottom: EK~Dra -- age $\sim 100$~Myr,
	 $P_{\rm rot} \sim 2.7$~d, $M_{\star} \sim 0.9~M_{\odot}$
	 (ObsID: 1884; 66~ks); 
	 $\pi^1$~UMa -- age $\sim 300$~Myr, $P_{\rm rot} \sim 4.7$~d,
	 $M_{\star} \sim 1~M_{\odot}$ (ObsID: 23; 30~ks);
	 $\alpha$~Cen -- age $\sim 6$~Gyr, $P_{\rm rot} \sim P_{\rm \odot}$ 
	 (i.e., $\sim 27$~d), $M_{\star} \sim 1.1~M_{\odot}$ 
	 (ObsID: 7432; 117~ks).
	 In the upper panel some of the strongest lines are labeled.
	 The intensity of \oviii\ emission relative to \ovii\ provides a 
	 visual indication of the temperature of the X-ray emitting plasma, 
	 being larger for higher temperatures.
	 Younger solar-like stars are characterized by higher 
	 X-ray emission levels ($L_{\rm X} \gtrsim 10^{30}$~erg~$s^{-1}$,
	 i.e.\ $\gtrsim 10^3 (L_{\rm X})_{\odot}$),
	 coronal temperatures ($T \gtrsim 10$~MK), and flaring rates.
	 These all decrease with age because of the reduced
	 efficiency of the underlying dynamo mechanism at the 
	 lower rotation rates due to substantial angular momentum 
	 loss.
	 }\label{fig:Sunlike}
\end{figure*}

Studies of large samples of solar-like stars at different evolutionary 
stages help investigate the evolution of the dynamo processes that
are mainly responsible for the X-ray production in these cool stars. 
In particular, studies of this type carried out with high resolution 
spectroscopy, while requiring a large investment of time and therefore 
focusing necessarily on small samples of stars, have nonetheless provided 
very important insights into the response of the corona to the decline in 
rotation-powered magnetic field generation and dissipation, and 
provide details of how X-ray emission on the Sun has evolved over 
time, as shown for instance by Telleschi et al.\ \cite{Telleschi05}.
This in turn could be relevant to the evolution of the solar system 
and the earth's atmosphere (see Feigelson's paper in this issue).
Within relatively short timescales, during the post T Tauri through 
early main sequence phase, the efficient mass loss spins down the star
significantly. This affects the dynamo process because the stellar 
rotation rate is one of the most important parameters driving the
dynamo.
As a consequence, the X-ray activity decreases, with coronal temperature,
$L_{\rm X}$, and flare rate all decreasing, as shown in 
fig.~\ref{fig:Sunlike} for three solar-like stars spanning ages
from $\sim 100$~Myr to $\sim 6$~Gyr.

\subsection{Element abundances}
\begin{figure}
\centerline{\includegraphics[width=10cm]{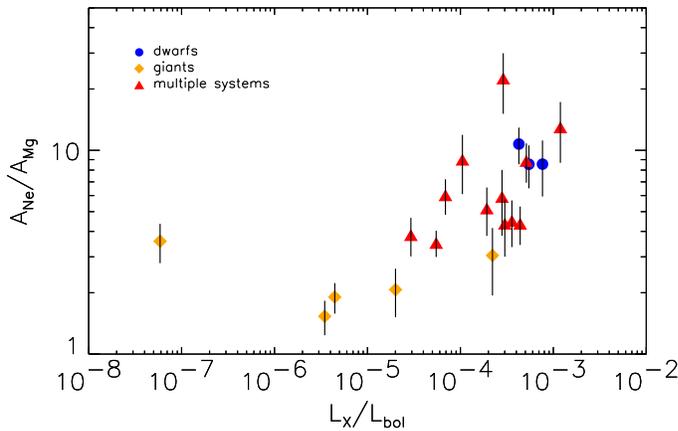}}
\caption{ Abundance ratio of high FIP Ne to low FIP Mg for 
	a sample of stars covering a wide range of activity.
	The abundance ratio is derived through a ratio of
	combination of H-like and He-like resonance lines,
	which is optimized to make the ratio largely temperature
	insensitive, as in \cite{DrakeTesta05}. The sample of
	spectra is the same analyzed by Drake \& Testa 
	\cite{DrakeTesta05}.
	 }\label{fig:abund}
\end{figure}

The study of element abundances has important implications 
in the wider astrophysical context and also for stellar physics.
For instance, chemical composition is a fundamental ingredient
for models of stellar structure since it significantly impacts
the opacity of the plasma.
Spectroscopic studies of the solar corona have provided a robust
body of evidence for element fractionation with respect to the 
photospheric composition (see e.g., \cite{Feldman92} and references
therein). Furthermore, this fractionation
effect appears to be a function of the element First Ionization
Potential (FIP), with low FIP elements such as Fe, Si, Mg, found to
be enhanced in the corona by a factor of a few, while high FIP elements
such as O have coronal abundances close to their photospheric values
(e.g., \cite{Feldman92}).  
This ``FIP effect'' has strong implications for the physical processes
at work in the solar atmosphere (see e.g., \cite{Laming04,Laming09} 
and references therein). 
Spectroscopic studies in the extreme ultraviolet have provided the 
first indication that in other stars as well the chemical composition of
coronal plasma is different from that of the underlying photosphere,
although with a dependence on FIP that is likely significantly different 
from that on the Sun (e.g., \cite{Drake96}).
High resolution X-ray spectroscopy with \cha\ and \xmm\ has for the 
first time provided robust and detailed information on the chemical
composition patterns of hot coronal plasma. 
Stellar coronae at the high end of the 
X-ray activity range appear characterized by an {\em inverse} FIP effect
(IFIP), i.e.\ with Fe significantly depleted in the corona, compared to 
the high FIP oxygen (e.g., \cite{Brinkman01}).
Investigation of element abundances in large samples of stars spanning
a large range of activity ($L_{\rm X}/L_{\rm bol} \sim 10^{-6}$--$10^{-3}$)
find a systematic gradual increase of IFIP effect with activity
level (e.g.,\cite{GAlvarez08}). This trend is shown in 
fig.~\ref{fig:abund} for the abundance ratio of low FIP element Mg to 
high FIP element Ne, derived from \cha\ \hetgs\ spectra for the same 
sample of stars for which Drake \& Testa studied the Ne/O abundance 
ratio \cite{DrakeTesta05}. 
An important caveat to keep in mind is that the stellar photospheric 
chemical composition is often unknown for the elements of interest, and the 
{\bf solar} photospheric composition is instead used as a reference 
\cite{Sanz04}.

In this context an interesting result is the behavior of Ne/O which 
remains rather constant over almost the whole observed range of activity
\cite{DrakeTesta05}, and, interestingly, this almost constant value is 
about 2.7 times higher than the adopted solar photospheric value. This 
might help to shed light on an outstanding puzzle in our understanding of 
our own Sun. Since Ne cannot be measured in the photosphere --
no photospheric Ne lines are present in the solar spectrum -- the 
solar photospheric Ne/O is not constrained. 
The remarkably constant Ne/O observed in stellar coronae, despite the 
significantly different properties of these stars, suggests that the 
observed coronal Ne/O 
actually reflects the underlying photospheric abundances. If the same
value is assumed for the solar photosphere as well, this would help 
resolve a troubling inconsistency between solar models and data from 
helioseismology observations \cite{AntiaBasu05}. It remains unresolved
though why the solar coronal Ne/O is found to be systematically lower 
than in other coronae (e.g., \cite{Young05}), though this is likely 
similar to other low activity stars \cite{Robrade08}. 
However, Laming \cite{Laming09} suggests that the low coronal Ne 
abundance on the Sun might be explained by the same fractionation 
processes that yield the general FIP effect.

\subsection{Spatial structuring of X-ray emitting plasma and dynamic events}
High spectral resolution in X-rays has made accessible a whole new
range of possible diagnostics for the spatial structuring of 
stellar coronae, for example: 

\begin{itemize}

\item {\bf opacity} effects in strong resonance lines yield estimates of 
path length, and therefore the spatial extent of X-ray emitting structures.
Only a handful of sources show scattering effects in their strongest
lines, and the derived lengths are very small when compared to the
stellar radii, analogous to solar coronal structures 
\cite{Testa04b,Matranga05,Testa07b}.

\item {\bf velocity modulation} derived from line shifts allows us to 
estimate the spatial distribution of the X-ray emitting plasma at different
temperatures, or the contribution of multiple system components to the 
total observed emission (e.g.,
\cite{Brickhouse01,Chung04,Ishibashi06,Huenemoerder06,Hussain07}). 
The unprecedented high spectral resolution of \cha\ is crucial for 
these studies with a velocity resolution down to $\sim 30$~km~s$^{-1}$ 
(e.g., \cite{Hoogerwerf04,Ishibashi06,Huenemoerder06}).

\item {\bf plasma density}, $n_e$, can be derived from the ratios of He-like 
triplets ($R = f/i \sim R_0 / [1 + n_e/n_c]$; \cite{GabrielJordan69}) 
\footnote{For cool stars the UV field is typically too weak to affect 
the He-like lines (which it does for hot stars as mentioned above) 
and therefore the $f/i$ ratio is mainly sensitive to the plasma density, 
above a critical density value which depends on the specific triplet 
(see \cite{GabrielJordan69}).}, therefore providing an estimate of the emitting
volumes, since the observed line intensity is proportional to $n_e^2 V$.
Several He-like triplet lines lie in the \cha\ and \xmm\ spectral range
covering a wide range of temperatures ($\sim 3-10$~MK from \ovii\ to \sixiii),
and densities ($\log (n_c[$cm$^{-3}]) \sim 10.5-13.5$ from \ovii\ to \sixiii).
We note that the unmatched resolving power of \cha\ \hetgs\ is crucial 
to resolve the numerous blends that affect the Ne and Mg triplets that cover
the important $\sim 3-6 \times 10^6$~K range.
Studies of plasma densities from He-like triplets in large samples of stars 
(\cite{Testa04a} studied \ovii, \mgxii, \sixiii, and \cite{Ness04} \ovii\ and
\neix) yield estimates of coronal filling factors which are remarkably small
especially for hotter plasma (typically $\ll 1$), but increase with 
X-ray surface flux \cite{Testa04a}.

\item {\bf flares} can provide clues on the size of the X-ray 
emitting structures and on the underlying physical processes that produce 
very dynamic events. 
The timescale of evolution of the flaring plasma (T, $n_e$) is related 
to the size of the flaring structure(s), and can be modeled to provide
constraints on the loop size (see e.g., \cite{Reale07} and references therein). 
Flares we observe in active stars involve much larger amounts of energy 
than observed on the Sun, with 
X-ray luminosities reaching values of $10^{32}$~erg~s$^{-1}$ and above,
i.e.\ more than two orders of magnitude larger than the most powerful 
solar flares. It is therefore not obvious that these powerful stellar 
flares are simply scaled up ($L_{\rm X}$, T, characteristic
timescales of evolution) versions of solar flares which we can study
and model with a much higher level of detail.
Novel diagnostics are provided by high resolution spectra,
and time-resolved high resolution spectroscopy of stellar flares is now 
possible with \cha\ and \xmm, at least for large flares in bright 
nearby sources.
G{\"u}del et al.\ \cite{Guedel02} have studied a large flare observed 
on Proxima Centauri, observing phenomena analogous to solar flaring
events: density enhancement during the flare, supporting the scenario 
of chromospheric evaporation, and the Neupert effect, i.e.\ proportionality
between soft X-ray emission and the integral of the non-thermal emission
(e.g., \cite{Hudson92}). \\
An interesting, and potentially powerful new diagnostic is provided by
{\bf Fe~K$\alpha$ }(6.4~keV, 1.94\AA) emission, which can be observed in
\cha\ and \xmm\ spectra. On the Sun Fe~K$\alpha$ emission has been 
observed during flares (e.g., \cite{Parmar84}) and it is interpreted 
as {\bf fluorescence} emission following inner shell ionization of 
{\em photospheric} neutral Fe due to hard X-ray coronal emission 
($> 7.11$~keV). In this scenario, the efficiency of Fe~K$\alpha$ 
production depends on the geometry, i.e.\ on the height of the 
source of hard ionizing continuum, through the dependence on the 
solid angle subtended and the average depth of formation of 
Fe~K$\alpha$ photons (e.g., \cite{Bai79,Drake08}).
In cool stars other than the Sun, Fe~K$\alpha$ has now been detected in
young stars with disks (see next section) where the fluorescent emission 
is thought to come from the cold disk material, and in only two,
supposedly diskless, sources during large flares: the G1 yellow giant 
HR~9024 \cite{Testa08a}, and the RS~CVn system II~Peg 
\cite{Osten07,Ercolano08}.
For HR~9024 the \cha\ \hetgs\ observations can be matched in detail
with a hydrodynamic model of a flaring loop yielding an estimate for
the loop height $h \sim 0.3 R_{\star}$ \cite{Testa07a}, and an 
{\em effective height} for the fluorescence production of 
$\sim 0.1 R_{\star}$ ($R_{\star}$ being the stellar radius). 
These values compare well with the value derived 
from the analysis of the measured Fe~K$\alpha$ emission, 
$h \lesssim 0.3 R_{\star}$.

\end{itemize}

\section{Young stars: powerful coronae, accretion, jets, magnetic fields and winds}
X-ray emission from young stars is presently one of the hot topics 
in X-ray astrophysics. Stellar X-rays are thought to significantly affect  
the dynamics, heating and chemistry of protoplanetary disks, influencing 
their evolution (see article by E.Feigelson in this same issue). 
Also, irradiation of close-in planets increases their mass loss rates 
possibly to the extent of complete evaporation of their atmospheres
(e.g., \cite{Penz08}).

Young stars are typically characterized by strong and variable X-ray
emission (e.g., \cite{Preibisch05}), and many recent \cha\ and \xmm\ 
studies have been investigating whether their coronae might just be 
powered up versions of their evolved main sequence counterparts, or
whether other processes might be at work in these early evolutionary stages.
For example, the observations have addressed the issue of accretion-related 
X-ray emission processes in accreting (classical) T Tauri stars (CTTS), on 
which material from a circumstellar disk is channeled onto the central
star by its magnetic field.

CTTS have observed X-ray luminosities that are systematically smaller 
by about a factor 2 than non accreting TTS (WTTS), (e.g., 
\cite{Preibisch05}). It is not yet clear however if accretion might 
suppress or obscure coronal X-rays, or instead, whether higher X-ray 
emission levels might increase photoevaporation 
of the accreting material, modulating the accretion rate \cite{Drake09}.

\subsection{Accretion related X-ray production}
\begin{figure}
\centerline{\includegraphics[width=8cm]{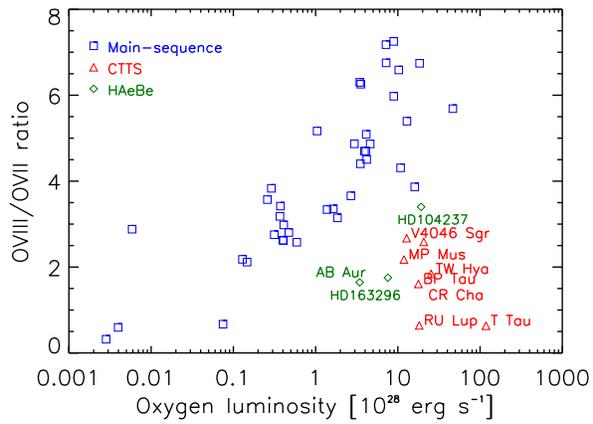}}
\caption{ The ratio of \oviii/\ovii\ vs.\ oxygen luminosity for 
	a large sample of main sequence and pre-main sequence stars
	shows the soft excess in high-resolution spectra of CTTS
	and HAe stars with respect to main sequence and non accreting
	stars. The figure is a modified version of fig.7 of
	G{\"u}nther et al.\ (\cite{Gunther09}), where the data point for 
	HD~104237 (using measured fluxes from \cite{Testa08}) has been added.
	 }\label{fig:HAe}
\end{figure}

High resolution spectroscopy has proved crucial for probing the physics
of X-ray emission processes in young stars. The first high resolution X-ray
spectrum of an accreting TTS, TW~Hya, has revealed obvious peculiarities
\cite{Kastner02} with respect to the coronal spectra of main sequence cool stars:
\begin{itemize}
\item {\bf very soft emission}: the X-ray spectrum of TW~Hya is 
	characterized by a temperature of only few MK ($\sim 3$~MK) 
	whereas coronae with comparable X-ray luminosities 
	($L_{\rm X} \sim 10^{30}$~erg~s$^{-1}$) typically have 
	strong emission at temperatures $\gtrsim$ 10~MK.
\item {\bf high $n_e$}: the strong cool He-like triplets of Ne and O
	have line ratios that imply very high densities 
	($n_e \gtrsim 10^{12}$~cm$^{-3}$), whereas in
	non-accreting sources typical densities are about two
	orders of magnitude lower.
\item {\bf abundance anomalies}: the X-ray spectrum of TW~Hya is 
	characterized by very low metal abundances, while Ne is 
	extremely high \cite{Stelzer04,Drake05} when compared to 
	other stellar coronae.
\end{itemize}
These peculiar properties strongly suggest that the X-ray emission of
TW~Hya is originating from shocked accreting plasma. Indeed, the observed
X-ray spectra of some of these sources have been successfully modeled 
as accretion shocks \cite{Gunther07,Sacco08}.
High resolution spectra subsequently obtained for other CTTS
have confirmed unusually high $n_e$ from the \ovii\ 
lines \cite{Schmitt05, Gunther06,Argiroffi07,Robrade07}, indicating 
that in these stars at least some of the observed X-rays are most 
likely produced through accretion-related mechanisms.
We note that TW~Hya is the CTTS for which the cool X-ray emission 
produced in the accretion shocks is the most prominent with respect
to the coronal emission, while all other CTTS for 
which high resolution spectra have been obtained have a much stronger
coronal component. For these latter sources we are able to probe 
accretion related X-rays only thanks to the high spectral resolution
which allows us to separate the two components.
Recent studies of optical depth effect in strong resonance lines in
CTTS provide confirmation of the high densities derived from the
He-like diagnostics \cite{Argiroffi09}.
Another diagnostic of accretion related X-ray production mechanisms is
offered by the \oviii/\ovii\ ratio which, in accreting TTS, is much larger
than in non-accreting TTS or main sequence stars
\cite{Guedel07} (see also fig.\ref{fig:HAe}).
Herbig~AeBe stars, young intermediate mass analogs of TTS, appear to 
share the same properties \cite{Robrade07}.

\subsection{Flaring activity and coronal geometry} 
X-ray emission of young stars is characterized by very high levels of 
X-ray variability pointing to very intense flaring activity in the
young coronae of TTS. This is beautifully demonstrated by 
the \cha\ Orion Ultradeep Project (COUP) of almost uninterrupted 
(spanning about 13~days) observations of the Orion Nebula Cluster star 
forming region\footnote{Movies of this dataset are available at 
http://www.astro.psu.edu/coup/.}.
Hydrodynamic modeling of some of the largest flares of TTS imply,
for some of these sources, very large sizes for the flaring structures 
($L \gtrsim 10 R_{\star}$). This may provide evidence of a star-disk 
connection \cite{Favata05}.
However, follow-up studies of these flares indicate that the largest 
structures seem to be associated with non-accreting sources, consistent
with the idea that in accreting sources, the inner disk, reaching close 
to the star, might truncate the otherwise very large coronal 
structures \cite{Getman08}.
In a few of these sources, with strong hard X-ray spectra, Fe~K$\alpha$ 
emission has been observed (see e.g., \cite{Tsujimoto05} for a survey of 
Orion stars).  
The Fe~K$\alpha$ emission is generally interpreted as fluorescence from
the circumstellar disk, however in a few cases the observed equivalent widths
are extremely high and apparently incompatible with fluorescence models 
(see e.g., \cite{Czesla07,Giardino07}). This apparent discrepancy could either 
be due to partial obscuration of the X-ray emission of the flare \cite{Drake08}
or could instead point to different physical processes at work, 
for instance impact excitation \cite{Emslie86}.

\subsection{Herbig Ae stars} 
In their pre-main sequence phase, intermediate mass stars appear to 
be moderate X-ray sources (e.g., \cite{Stelzer09} and references therein).
Their X-ray emission characteristics are overall similar to the lower 
mass TTS (hot, variable), possibly implying that the same X-ray emission 
processes are at work in the two classes of stars, or that the emission
is due to unseen TTS companions.
However, a handful of Herbig Ae stars show unusually soft X-ray emission:
e.g., AB~Aur \cite{Telleschi07}, HD~163296 \cite{Swartz05,Gunther09}.
High resolution spectra have been obtained for these stars, together
with HD~104237 \cite{Testa08}. 
One similarity with the high resolution spectra of CTTS,
appears to be the presence of a soft excess (\oviii/\ovii), compared to 
coronal sources, as shown by  \cite{Gunther09} (see figure~\ref{fig:HAe}).
However their He-like triplets are generally compatible with
low density, at odds with the accreting TTS (with maybe the exception
of HD~104237, possibly indicating higher $n_e$).
AB~Aur and HD~104237 have X-ray emission that seems to be modulated on 
timescales comparable with the rotation period of the A-type star 
therefore rendering the hypothesis that X-ray emission originates from
low-mass companions less plausible.

\section{Conclusions}
The past decade of stellar observations has led to exciting progress
in our understanding of the X-ray emission processes in stars, also 
shifting in the process the perspective of stellar studies which are 
now much more focused on star and planet formation.
In particular high resolution X-ray spectroscopy, available for the first 
time with \cha\ and \xmm, is now playing a crucial role in constraining 
and developing models of X-ray emission, e.g., for early-type stars, 
late-type stellar coronae, and in the case of young stars, by providing 
a unique means for probing accretion related X-ray emission processes, 
as well as the opportunity to examine the effects of X-rays on the 
circumstellar environment.

\subsection{Progress and some open issues}
X-ray emission processes in early-type stars now present a much more
complex scenario, in which magnetic fields also likely play a key role.
Some puzzling results found for several massive stars concern the hard, 
variable X-ray spectra with relatively narrow lines, which 
cannot be explained by existing models. 

Spectroscopic studies of large samples of stars have provided robust 
findings on chemical fractionation in X-ray emitting plasma, which 
now require improved models to understand the physical processes
yielding the observed abundance anomalies.

A satisfactory understanding of activity cycles is lacking even 
for our own Sun, and recent discoveries of X-ray cycles on other stars
can provide further constrains for dynamo models.

We are now taking the first steps in studying flares with temporally
resolved high resolution spectra, and this will greatly help
constrain our models and really test whether the physics of these
dynamic events, in the extreme conditions seen in some cases
(e.g., $T \gtrsim 10^8$~K), are still the same as for
solar flares. At present, the effective areas are often 
insufficient to obtain good S/N at high spectral resolution 
on the typical timescales of the plasma evolution
during these very dynamic events. The International X-ray Observatory
(IXO), in the planning stages for a launch about a decade from now, 
will make a large number of stars accessible for this kind of study.

In young stars, a very wide range of phenomena are observed to occur,  
and while this young field has already offered real breakthroughs
there is still a long way to go to understand the details of 
accretion, jets, extremely large X-ray emitting structures, 
the influence of X-rays on disks and planets, and the
interplay between accretion and X-ray activity.





\begin{acknowledgments}
This work has greatly benefited greatly from discussions with several people,
and, in particular, I would like to warmly thank Jeremy Drake and Manuel G{\"u}del.
I also would like to thank Hans Moritz G{\"u}nther for permission to use
original figure material. This work has been supported by NASA grant GO7-8016C.
\end{acknowledgments}


\begin{thebibliography}{}
\bibitem{Friedman51} Friedman H., Lichtman S.W.\ \& Byram E.T.\ (1951)
	Photon counter measurements of solar X-rays and Extreme 
	Ultraviolet light. Ph.\ Rv. 83:1025-1030.
\bibitem{Vaiana81} Vaiana, G.S.\ et al.\ (1981) Results from an 
	extensive Einstein stellar survey. ApJ 245:163-182.
\bibitem{LucyWhite80} Lucy L.B.\ \& White R.L.\ (1980) X-ray emission 
	from the winds of hot stars. ApJ 241:300-305.
\bibitem{Owocki88} Owocki S.P., Castor J.I.\ \& Rybicki G.B.\ (1988)
	Time-dependent models of radiatively driven stellar winds. 
	I - Nonlinear evolution of instabilities for a pure absorption 
	model. ApJ 335:914-930.
\bibitem{Czesla07a} Czesla S., \& Schmitt J.H.H.M.\ (2007) Are magnetic 
	hot stars intrinsic X-ray sources? A\&A 465:493-499.
\bibitem{Schroeder07} Schr{\" o}der C.\ \& Schmitt J.H.M.M.\ (2007) 
	X-ray emission from A-type stars. A\&A 475:677-684.
\bibitem{Canizares05} Canizares C.R.\ et al.\ (2005) The Chandra High-Energy 
	Transmission Grating: Design, Fabrication, Ground Calibration, 
	and 5 Years in Flight. PASP 117:1144-1171.
\bibitem{Donati02} Donati J.F.\ et al.\ (2002) The magnetic field and 
	wind confinement of {$\theta^{1}$} Orionis C. MNRAS 333:55-70.
\bibitem{Cassinelli01} Cassinelli J.P., Miller N.A., Waldron W.L., 
	MacFarlane J.J.\ \& Cohen D.H.\ (2001) Chandra Detection of 
	Doppler-shifted X-Ray Line Profiles from the Wind of {$\zeta$} 
	Puppis (O4 F). ApJL 554:55-58.
\bibitem{Cohen06} Cohen D.H.\ et al.\ (2006) Wind signatures in the X-ray 
	emission-line profiles of the late-O supergiant {$\zeta$} Orionis.
	MNRAS 368:1905-1916.
\bibitem{Miller02} Miller N.A., Cassinelli J.P., Waldron W.L., MacFarlane 
	J.J.\ \& Cohen D.H.\ (2002) New challenges for wind shock 
	models: the Chandra spectrum of the hot star {$\delta$} Orionis.
	 ApJ 577:951-960.
\bibitem{Skinner08} Skinner S.L\ et al.\ (2008) High-resolution 
	Chandra X-ray imaging and spectroscopy of the {$\sigma$} 
	Orionis cluster. ApJ 683:796-812.
\bibitem{Gagne05} Gagn{\'e} M.\ et al.\ (2005) Chandra HETGS 
	multiphase spectroscopy of the young magnetic O star 
	{$\theta^1$} Orionis C. ApJ 628:986-1005.
\bibitem{Cohen03} Cohen D.H.\ et al.\ (2003) High-Resolution Chandra 
	Spectroscopy of {$\tau$} Scorpii: A Narrow-Line X-Ray Spectrum 
	from a Hot Star. ApJ 586:495-505.
\bibitem{Leutenegger06} Leutenegger M.A., Paerels F.B.S., Kahn 
	S.M.\ \& Cohen D.H.\ (2006) Measurements and analysis of 
	Helium-like triplet ratios in the X-ray spectra of O-type 
	stars. ApJ 650:1096-1110.
\bibitem{Owocki01} Owocki S.P.\ \& Cohen D.H.\ (2001) 
	X-ray line profiles from parameterized emission within an 
	accelerating stellar wind. ApJ 559:1108-1116.
\bibitem{Favata08} Favata F., Micela G., Orlando S., Schmitt J.H.M.M.\ 
	\& Sciortino S.\ (2008) The X-ray cycle in the solar-type star 
	HD 81809. XMM-Newton observations and implications for the 
	coronal structure. A\&A 490:1121-1126.
\bibitem{Hempelmann06} Hempelmann A.\ et al.\ (2006) Coronal activity 
	cycles in 61 Cygni. A\&A 460:261-267.
\bibitem{Ayres09} Ayres T.R.\ (2009) The Cycles of {$\alpha$} Centauri. 
	ApJ 696:1931-1949.
\bibitem{Telleschi05} Telleschi A.\ et al.\ (2005) 
	Coronal evolution of the Sun in time: high-resolution X-ray 
	spectroscopy of solar analogs with different ages. ApJ 622:653-679.
\bibitem{Feldman92} Feldman U.\ (1992) Elemental abundances in the 
	upper solar atmosphere. Phys.\ Scr. 46:3:202-220.
\bibitem{Laming04} Laming J.M.\ (2004) A unified picture of the 
	first ionization potential and inverse first ionization 
	potential effects. ApJ 614:1063-1072.
\bibitem{Laming09} Laming J.M.\ (2009) Non-Wkb models of the first 
	ionization potential effect: implications for solar coronal 
	heating and the coronal Helium and Neon abundances. 
	ApJ 695:954-969.
\bibitem{Drake96} Drake J.J., Laming J.M.\ \& Widing K.G.\ (1996) The FIP 
	effect and abundance anomalies in late-type stellar coronae.
	IAU Colloq.\ 152: Astrophysics in the Extreme Ultraviolet, 
	Ed.\ S. Bowyer \& R.F.\ Malina, 97.
\bibitem{Brinkman01} Brinkman A.C.\ et al.\ (2001) First light measurements 
	with the XMM-Newton reflection grating spectrometers: Evidence 
	for an inverse first ionisation potential effect and anomalous 
	Ne abundance in the Coronae of HR 1099. ApJL 365:324-328.
\bibitem{GAlvarez08} Garc{\'i}a-Alvarez D., Drake J.J., Kashyap V.L.,
	Lin L.\ \& Ball B.\ (2008) Coronae of young fast rotators.
	ApJ 679:1509-1521.
\bibitem{DrakeTesta05} Drake J.J.\ \& Testa P.\ (2005) The `solar model 
	problem' solved by the abundance of neon in nearby stars. 
	Nature 436:525-528.
\bibitem{Sanz04} Sanz-Forcada J., Favata F.\ \& Micela G.\ (2004)
	 Coronal versus photospheric abundances of stars with 
	 different activity levels. A\&A 416:281-290.
\bibitem{AntiaBasu05} Antia H.M.\ \& Basu S.\ (2005) 
	The discrepancy between solar abundances and helioseismology. 
	ApJL 620:129-132.
\bibitem{Young05} Young, P.R.\ (2005) The Ne/O abundance ratio 
	in the quiet Sun. A\&A 444:L45-L48.
\bibitem{Robrade08} Robrade J., Schmitt J.H.M.M.\ \& Favata F.\ (2008) 
	Neon and oxygen in low activity stars: towards a coronal 
	unification with the Sun. A\&A 486:995-1002.
\bibitem{Testa04b} Testa P., Drake J.J., Peres G.\ \& DeLuca E.E.\ (2004)
  	Detection of X-ray resonance scattering in active stellar 
	coronae. ApJL 609:L79-L82.
\bibitem{Matranga05} Matranga M., Mathioudakis M., Kay H.R.M.\ \& 
	Keenan F.P.\ (2005) Flare X-ray observations of AB Doradus: 
	evidence of stellar coronal opacity. ApJL 621:L125-L128.
\bibitem{Testa07b} Testa P., Drake J.J., Peres G.\ \& Huenemoerder 
	D.P.\ (2007) On X-ray optical depth in the coronae of active 
	stars. ApJ 665:1349-1360.
\bibitem{Brickhouse01} Brickhouse N.S., Dupree A.K\ \& Young P.R.\ (2001)
	X-Ray Doppler Imaging of 44i Bootis with Chandra. ApJL 562:75-78.
\bibitem{Chung04} Chung S.M., Drake J.J., Kashyap V.L., Lin L.\
	 \& Ratzlaff P.W.\ (2004) Doppler Shifts and Broadening and the 
	 Structure of the X-Ray Emission from Algol. ApJ 606:1184-1195.
\bibitem{Ishibashi06} Ishibashi K., Dewey D., Huenemoerder D.P.\ \& 
	Testa, P.\ (2006) Chandra/HETGS observations of the Capella 
	system: the primary as a dominating X-ray source. 
	ApJL 644:L1171-L1120.
\bibitem{Huenemoerder06} Huenemoerder D.P., Testa P.\ \& Buzasi D.L.\ 
	(2006) X-ray spectroscopy of the Contact binary VW Cephei.
	ApJ 650:1119-1132.
\bibitem{Hussain07} Hussain, G.A.J.\ et al.\ (2007) The coronal 
	structure of AB Doradus determined from contemporaneous 
	Doppler imaging and X-ray spectroscopy. MNRAS 377:1488-1502.
\bibitem{Hoogerwerf04} Hoogerwerf R., Brickhouse N.S.\ \& Mauche 
	C.W.\ (2004) The radial velocity and mass of the white dwarf 
	of EX Hydrae measured with Chandra. ApJ 610:411-415.
\bibitem{GabrielJordan69} Gabriel A.H. \& Jordan C.\ (1969) 
	Interpretation of solar helium-like ion line intensities.
	MNRAS 145:241.
\bibitem{Testa04a} Testa P., Drake J.J.\ \& Peres G.\ (2004) 
	The density of coronal plasma in active stellar coronae. 
	ApJ 617:508-530.
\bibitem{Ness04} Ness J.-U., G{\"u}del M., Schmitt J.H.M.M., 
	Audard M.\ \& Telleschi A.\ (2004) On the sizes of stellar 
	X-ray coronae. A\&A 427:667-683.
\bibitem{Reale07} Reale F.\ (2007) Diagnostics of stellar flares from 
	X-ray observations: from the decay to the rise phase. 
	A\&A 417:271-279.
\bibitem{Guedel02} G{\" u}del M., Audard M., Skinner S.L.\ \& Horvath
	M.I.\ (2002) X-ray evidence for flare density variations 
	and continual chromospheric evaporation in Proxima Centauri.
	ApJL 580:L73-L76.
\bibitem{Hudson92} Hudson H.S., Acton L.W., Hirayama T.\ \& 
	Uchida Y.\ (1992) White-light flares observed by YOHKOH. 
	PASJ 44:L77-L81.
\bibitem{Parmar84} Parmar A.N.\ et al.\ (1984) SMM observations of 
	K-{$\alpha$} radiation from fluorescence of photospheric iron 
	by solar flare X-rays. ApJ 279:866-874.
\bibitem{Bai79} Bai T.\ (1979) Iron K-{$\alpha$} fluorescence in solar 
	flares - A probe of the photospheric iron abundance.  
	Sol.\ Phys. 62:113-121.
\bibitem{Drake08} Drake J.J., Ercolano B.\ \& Swartz D.A.\ (2008) 
	X-Ray-fluorescent Fe K{$\alpha$} lines from stellar 
	photospheres. ApJ 678:385-393.
\bibitem{Testa08a} Testa P.\ et al.\ (2008) Geometry diagnostics of 
	a stellar flare from fluorescent X-rays. ApJL 675:L97-L100.
\bibitem{Osten07} Osten R.A.\ et al.\ (2007) Nonthermal hard X-ray 
	emission and iron K{$\alpha$} emission from a superflare on 
	II Pegasi. ApJ 654:1052-1067.
\bibitem{Ercolano08} Ercolano B., Drake J.J., Reale F., Testa 
	P.\ \& Miller J.M.\ (2008) Fe K{$\alpha$} and hydrodynamic loop 
	model diagnostics for a large flare on II Pegasi. ApJ 688:1315-1319.
\bibitem{Testa07a} Testa P., Reale F., Garcia-Alvarez D., \& 
	Huenemoerder D.P.\ (2007) Detailed diagnostics of an X-ray 
	flare in the single giant HR 9024. ApJ 663:1232-1243.
\bibitem{Penz08} Penz T., Micela G.\ \& Lammer H.\ (2008)
	Influence of evolving stellar X-ray luminosity distribution
	on exoplanetary mass loss. A\&A 477:309-314.
\bibitem{Preibisch05} Preibisch T.\ et al.\ (2005) The origin of T 
	Tauri X-ray emission: new insights from the Chandra Orion 
	Ultradeep Project. ApJS 160:401-422.
\bibitem{Drake09} Drake J.J., Ercolano B., Flaccomio E., \& Micela G.\ (2009) 
	X-ray photoevaporation-starved T Tauri accretion. ApJL 699:35-38.
\bibitem{Kastner02} Kastner J.H., Huenemoerder D.P., Schulz N.S., 
	Canizares C.R.\ \& Weintraub D.A.\ (2002) Evidence for 
	accretion: high-resolution X-ray spectroscopy of the 
	classical T Tauri star TW Hydrae. ApJ 567:434-440.
\bibitem{Stelzer04} Stelzer B.\ \& Schmitt J.H.M.M.\ (2004) 
	X-ray emission from a metal depleted accretion shock 
	onto the classical T Tauri star TW Hya. A\&A 418:687-697.
\bibitem{Drake05} Drake J.J., Testa P.\ \& Hartmann L.\ (2005) X-Ray 
	diagnostics of grain depletion in matter accreting onto 
	T Tauri stars. ApJL 627:149-152.
\bibitem{Gunther07} G{\"u}nther H.M., Schmitt J.H.M.M.,
	Robrade J.\ \& Liefke C.\ (2007) X-ray emission from classical 
	T Tauri stars: accretion shocks and coronae?. A\&A 466:1111-1121.
\bibitem{Sacco08} Sacco G.G.\ et al.\ (2008) X-ray emission from dense 
	plasma in classical T Tauri stars: hydrodynamic modeling of 
	the accretion shock. A\&A 491:L17-L20.
\bibitem{Schmitt05} Schmitt J.H.M.M., Robrade J., Ness J.U., Favata
	F.\ \& Stelzer B.\ (2005) X-rays from accretion shocks in T 
	Tauri stars: The case of BP Tau. A\&A 432:L35-L38.
\bibitem{Gunther06} G{\"u}nther H.M., Liefke C., Schmitt J.H.M.M.,
	Robrade J.\ \& Ness J.U.\ (2006) X-ray accretion signatures 
	in the close CTTS binary V4046 Sagittarii. A\&A 459:L29-L32.
\bibitem{Argiroffi07} Argiroffi C., Maggio A.\ \& Peres 
	G.\ (2007) X-ray emission from MP Muscae: an old classical 
	T Tauri star. A\&A 465:5-8.
\bibitem{Robrade07} Robrade J.\ \& Schmitt J.H.M.M.\ (2007) 
	X-rays from RU Lupi: accretion and winds in classical T Tauri 
	stars. A\&A 473:229-238.
\bibitem{Argiroffi09} Argiroffi C.\ et al.\ (2009) 
	X-ray optical depth diagnostics of T Tauri accretion shocks. 
	A\&A 507:939-948.
\bibitem{Guedel07} G{\" u}del M.\ \& Telleschi A.\ (2007)
	 The X-ray soft excess in classical T Tauri stars. A\&A 474:L25-L28.
\bibitem{Favata05} Favata F.\ et al.\ (2005) Bright X-Ray flares in Orion young 
	stars from COUP: evidence for star-disk magnetic fields?. ApJS 160:469-502.
\bibitem{Getman08} Getman K.V.\ et al.\ (2008) X-ray flares 
	in Orion young stars. II. Flares, magnetospheres, and 
	protoplanetary disks. ApJ 688:437-455.
\bibitem{Tsujimoto05} Tsujimoto M.\ et al.\ (2005) Iron fluorescent 
	line emission from young stellar objects in the Orion 
	Nebula. ApJS 160:503-510.
\bibitem{Czesla07} Czesla S., \& Schmitt J.H.H.M.\ (2007) The nature of 
	the fluorescent iron line in V 1486 Orionis. A\&A 470:L13-L16.
\bibitem{Giardino07} Giardino G.\ et al.\ (2007) 
	Results from Droxo. I. The variability of fluorescent Fe 
	6.4~keV emission in the young star Elias 29. 
	High-energy electrons in the star's accretion tubes?.
	A\&A 475:891-900.
\bibitem{Emslie86} Emslie A.G., Phillips K.J.H.\ \& Dennis B.R.\ (1986)
	The excitation of the iron K-{$\alpha$} feature in solar flares.
	Sol.\ Phys.\ 103:89-102.
\bibitem{Stelzer09} Stelzer B., Robrade J., Schmitt J.H.M.M.\ \& 
	Bouvier J.\ (2009) New X-ray detections of Herbig stars. 
	A\&A 493:1109-1119.
\bibitem{Telleschi07} Telleschi A.\ et al.\ (2007) 
	The first high-resolution X-ray spectrum of a Herbig star: 
	AB Aurigae. A\&A 468:541-556.
\bibitem{Swartz05} Swartz D.A.\ et al.\ (2005) The Herbig Ae star 
	HD 163296 in X-rays. ApJ 628:811-816.
\bibitem{Gunther09} G{\"u}nther H.M.\ \& Schmitt J.H.M.M.\ (2009) 
	The enigmatic X-rays from the Herbig star HD 163296: 
	Jet, accretion, or corona?. A\&A 494:1041-1051.
\bibitem{Testa08} Testa P., Huenemoerder D.P., Schulz N.S.\ \&
	Ishibashi K.\ (2008) X-Ray emission from young stellar objects 
	in the {$\epsilon$} Chamaeleontis group: the Herbig Ae star 
	HD 104237 and associated low-mass stars. ApJ 687:579-597.
\end{thebibliography}




\end{article}






\end{document}